# Structural and Magnetic Characterization of $Cu_xMn_{1-x}Fe_2O_4$ (*x= 0.0, 0.25*) Ferrites Using Neutron Diffraction and Other Techniques

I.B. Elius[1†], A.K.M. Zakaria[1,5], J. Maudood[1], S. Hossain[1], M.M. Islam[2], A. Nahar[3], Md Sazzad Hossain[4],

I. Kamal[5]

[1]Institute of Nuclear Science and Technology, Bangladesh Atomic Energy Commission, GPO Box No.3787, Dhaka-1000 Bangladesh
[2]Department of Applied Chemistry and Chemical Engineering, University of Dhaka, Dhaka-1000
[3]Materials Science Division, Atomic Energy Centre, Dhaka-1000
[4] Department of Physics, University of Dhaka, Dhaka-1000
[5]Bangladesh Atomic Energy Commission, Agargaon, Dhaka-1207

[†]Corresponding author: iftakhar.elius@gmail.com

**ABATRACT**

**Manganese ferrite ($MnFe_2O_4$) and copper doped manganese ferrite ($Mn_{0.75}Cu_{0.25}Fe_2O_4$) soft materials were synthesized through solid-state sintering method. The phase purity and quality were confirmed from x-ray diffraction patterns. Then the samples were subjected to neutron diffraction experiment and the diffraction data were analyzed using FullProf software package. The surface morphology of the soft material samples was studied using a scanning electron microscope (SEM). Crystal parameters, crystallite parameters, occupancy at A and B sites of the spinel structure, magnetic moments of the atoms at various locations, symmetries, oxygen position parameters, bond lengths etc. were measured and compared with the reference data. In $MnFe_2O_4$, both octahedral (A) and tetrahedral (B) positions are shared by $Mn^{2+}$ and $Fe^{2+/3+}$ cations, here A site is predominantly occupied by $Fe^{2+}$ and B site is occupied by Mn at 0.825 occupancy. The $Cu^{2+}$ ions in $Cu_{0.25}Mn_{0.75}Fe_2O_4$ mostly occupy the B site. Copper mostly occupy the Octahedral (16d) sites. The length of the cubic lattice decreases with the increasing Copper content. The magnetic properties, i.e. A or B site magnetic moments, net magnetic moment etc. were measured using neutron diffraction analysis and compared with the bulk magnetic properties measured with VSM studies.**

## 1. Introduction

Manganese ferrites ($MnFe_2O_4$) are soft ferrite materials that have properties like high saturation magnetization, high permeability, low coercivity ($H_c$), low magnetostriction, low anisotropy, high curie temperature ($T_c$), high electrical resistivity etc. Thus soft magnetic materials have wide use in computer memory devices, microwave devices, RF-coil fabrication, antennas, transformer cores, magnetic recording media etc. [1] Since the last few decades, ferrites have also been used in wastewater treatment processes, carbon dioxide decomposition for the utilization of carbon as solar hydrogen carriers, hybridization process for mixing solar and fossil fuel sources, and conversion of solar energy into hydrogen energy[2]. Ferrites are also used in biomedical sector for drug delivery, to mark particular cells explicitly, peptide synthesis, for enhancement of X-ray imaging, MRI, sono-imaging, in tumor treatment etc.[3]

Since Bragg and Nishikawa determined the spinel structure in 1915 [4], undoubtedly, spinel ferrites are one of the most studied classes of materials because of its novel magnetic and electric properties. Spinel ferrites are in general represented by the formula $AB_2O_4$. In the formula $AB_2O_4$, '$A^{2+}$' are divalent cations (i.e. $Mn^{2+}$, $Zn^{2+}$, $Ni^{2+}$, $Fe^{2+}$ etc.) which occupy the tetrahedral interstitial positions. On the other hand, '$B^{3+}$' are trivalent cations (i.e. $Fe^{3+}$, $Mn^{3+}$, etc.) that occupy the octahedral sites. However, this general formula is not always necessarily followed; trivalent cations can take over tetragonal sites, and as a result, the divalent ions are made to occupy the octahedral sites improper fraction making the spinel structure fractionally or even fully inverted [5]. The cation concentrations of spinel ferrites depend upon factors like cations, their fractions, method of synthesis, which may eventually show different structural, magnetic, and electrical properties.

One of the essential factors that control different properties of ferrites is the structural arrangements of the cations. Hence, in this study, various structural parameters by Rietveld refinement of the XRD pattern and neutron diffraction analysis were determined and compared with the theoretical calculations. The parameters include the cell parameter, oxygen position parameter, space group, the bond lengths, and interaction angles. Moreover, the M-H curves were determined to study the impact of the Cu-doping in the samples.

## 2. Experimental

The polycrystalline samples of $MnFe_2O_4$ and $Mn_{0.75}Cu_{0.25}Fe_2O_4$ spinel oxides were prepared by conventional solid-state sintering method at the sample preparation laboratory of Institute of Nuclear Science and Technology, Atomic Energy Research Establishment (AERE), Bangladesh Atomic Energy Commission (BAEC). High purity oxides of $Fe_2O_3$ (99.00%), $Cu(II)O$ (99.00%), $MnO_2$ (99.00%) Sigma-Aldrich, UK, were mixed thoroughly by maintaining the stoichiometric proportions using an agate mortar and pestle for 2h. Then the mixtures were ground for 6 h in a stainless-steel ball mill. To ensure fine mixing, a small amount of distilled water was used as a milling fluid. Then the mixture was dried using a magnetic heater in the air until it turned into powder. The mixture was then calcined at 1000°C for 12 hours in the air. The temperature was raised at a rate of 2°C per minute and after heating the cooling rate was 4°C per minute. Then the mixtures were mixed again in an agate mortar and pressed into pellets using a hydraulic press.

Then the pellets were heated in a muffle furnace where the final firing temperature was 1050°C for $MnFe_2O_4$ and $Mn_{0.75}Cu_{0.25}Fe_2O_4$ were heated at a final temperature of 1100°C for 24 h. The temperature was raised at a rate of 2°C per minute and after heating, the cooling rate was 4°C per minute.

To determine the phase purity and quality, the samples were subjected to X-ray diffraction experiment at CARS, University of Dhaka with $Cu(K\alpha)$ radiation of wavelength 1.54178 Å in the span of $10° \leq 2\theta \leq 70°$ with a step size of 0.02°. Both the diffraction patterns exhibited sharp peaks corresponding to the characteristic peaks of cubic spinel structure.

Neutron diffraction analysis is a unique technique which can provide insights into the magnetic structure of the ferromagnetic, anti-ferromagnetic or paramagnetic materials. To investigate the magnetic structure along with the molecular structure, the samples were subjected to neutron diffraction experiment at Savar Neutron Diffractometer (SAND) which utilizes thermal neutron beam emerging from TRIGA Mark-II research reactor at Atomic Energy Research Establishment (AERE), Savar. The machine is a set of 15 $^3He_2$ gas-filled neutron detectors, which can maneuver around the sample table from $2\theta$ range of 5˚ to 115˚. The powder sample was kept in a Vanadium can (as Vanadium is transparent to thermal neutrons), on a rotating table. The neutron beam is monochromated using a set of silicon single crystal to a single wavelength beam of $\lambda$ = 1.5656 Å. The readings of the fifteen parallel detectors are then averaged and noise from stray radiations are suppressed at a data accusation unit[6].

After the neutron diffraction experiment, the diffraction data were refined using Reitveld based refinement code FullProf. The structural symmetry, cell parameters, atomic position parameters, magnetic moments of different ions at different positions were measured using this code.

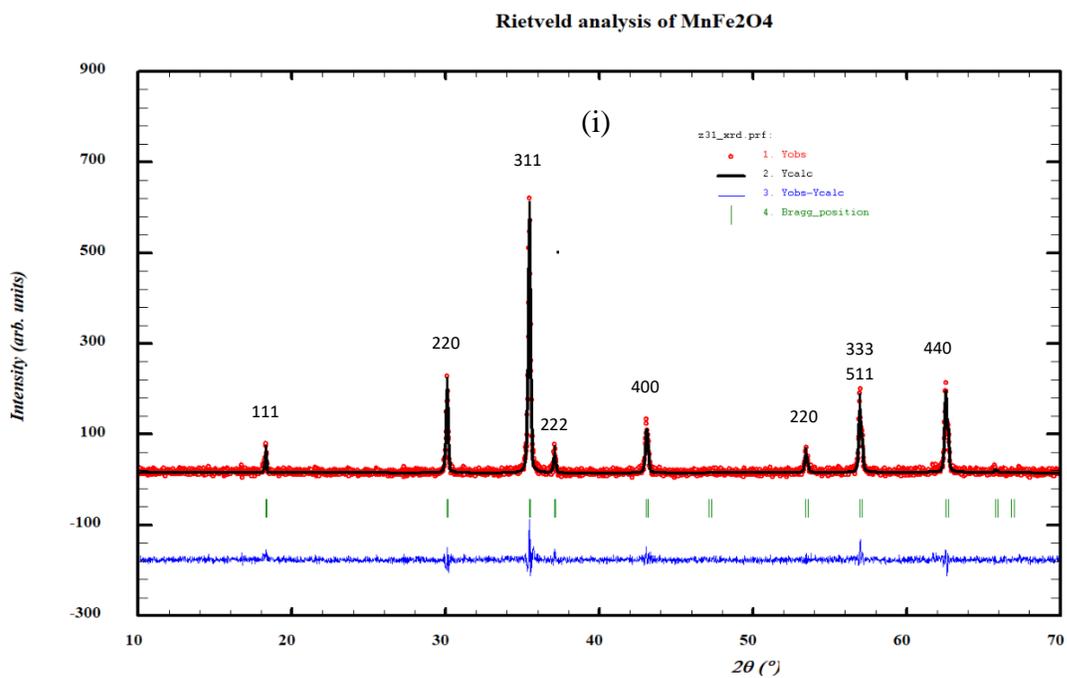

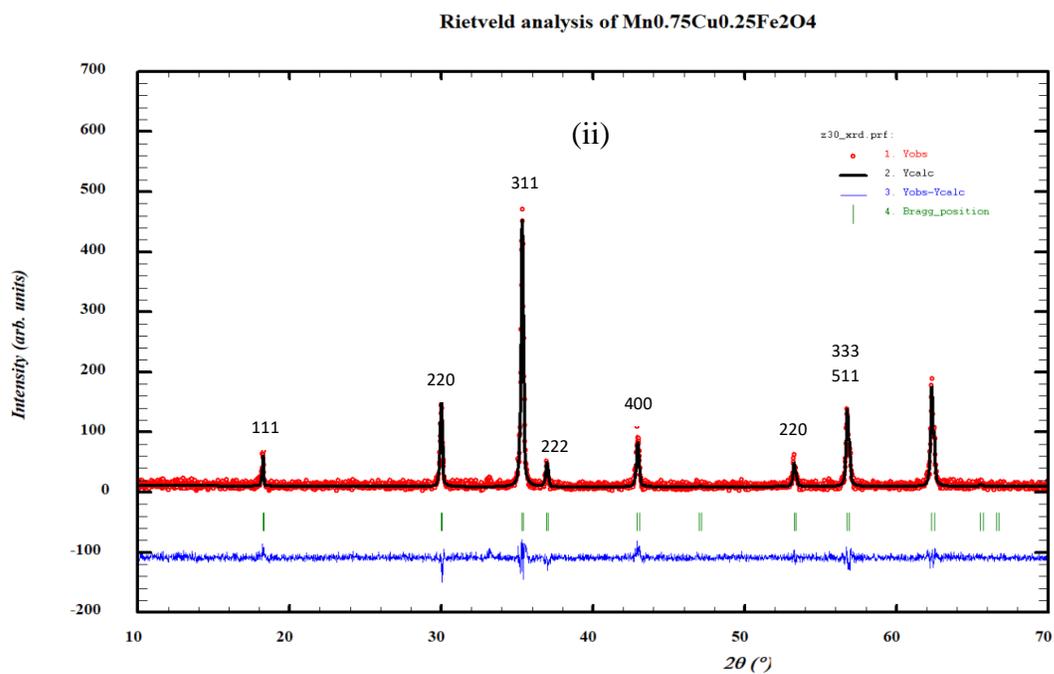

**Fig.1:** X-ray diffraction patterns of (i) $MnFe_2O_4$ and (ii) $Mn_{0.75}Cu_{0.25}Fe_2O_4$, fitted with FullProf refinement software. The red points are observed data points, the dark solid line indicates the refined model, the green markers are the peak locations and the blue points below are the difference curve between the $Y_{obs}$ and the $Y_{calc}$.

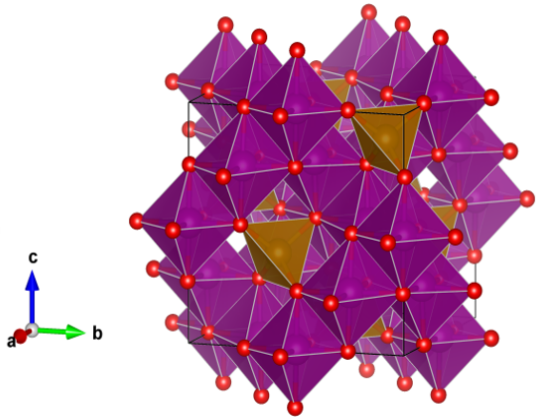

**Fig. 2:** Structure of MnFe$_2$O$_4$ crystal generated from the refinement of XRD data using VESTA software.

After that, Scanning Electron Microscopy (SEM) experiment were done using an FEI Inspect S50 to study the grain size distribution as well as the surface properties of the materials. The images were taken at 5000 times magnification applying 12.50 kV potential difference for MnFe$_2$O$_4$ and 20.00 kV for Cu$_{0.25}$Mn$_{0.75}$Fe$_2$O$_4$.

## 3. Results and Discussion

### 3.1 XRD Analysis

The X-ray diffraction patterns of the samples were yielded from a Rigaku X-ray diffractometer within a *2θ* range of 10° to 70°. The XRD patterns show prominent peaks at the locations identical to that of spinel ferrites having *Fd-3m* crystal symmetry. The absence of non-identical peaks indicates the phase purity of the synthesized samples. The indexing of the XRD peaks was done with the software '*Chekcell*', a small but elegant software that affirmed the assumed symmetry and provided a preliminary idea about the cell parameters[7]. The refined parameters are presented in table 1.

**Table 1:** The Rietveld refined parameters from XRD patterns.

| Sample | Symmetry | Cell parameter, a (Å) | Oxygen position parameter, u ($\bar{4}3m$) | X-ray density, d$_{xray}$ (gm/cc) |
|---|---|---|---|---|
| MnFe$_2$O$_4$ | Fd$\bar{3}$m | 8.377974 | 0.397837253 | 3.63826 |
| Mn$_{0.75}$Cu$_{0.25}$Fe$_2$O$_4$ |  | 8.402696 | 0.391561301 | 3.63985 |

### 3.2 Cation distribution

The cation distribution of the spinel ferrites is sensitive to the observed and calculated intensity ratios of peaks at (220), (400) and (440) planes. This method selects ratios of pairs of intensities,

$$\frac{I^{Obs}_{hkl}}{I^{Obs}_{h'k'l'}} = \frac{I^{Cal}_{hkl}}{I^{Cal}_{h'k'l'}} \quad (1)$$

where the relative integrated intensity of the XRD peaks can be calculated from the following formula

$$I_{hkl} = |F|^2 P L_p \quad (2)$$

where F is the structure factor, P is the multiplicity factor and *L$_p$* is the Lorentz polarization factor which in fact depends solely on Bragg's angle *2θ,* as stated below[8]

$$L_p = \frac{1+ cos^2 2\theta}{sin^2 2\theta cos 2\theta} \quad (3)$$

$$R = \left| \frac{I_{hkl}^{Obs}}{I_{h'k'l'}^{Obs}} - \frac{I_{hkl}^{Cal}}{I_{h'k'l'}^{Cal}} \right| \quad (4)$$

Considering all possible cation distributions of $Mn^{2+}$, $Fe^{2+}$, $Fe^{3+}$, $Cu^{2+}$ at tetrahedral and octahedral positions both the composition minimizing the agreement factor (***R***), yield the correct cation distribution.

The cation distribution is presented below in table 2.

**Table 2:** The crystallographic parameters, peak positions, half-width parameters, along with the observed and calculated intensities of the major peaks.

| Sample | obs | h | k | l | Multiplicity | $d_{hkl}$ | Half-width parameter (W) | Observed intensity ($I_{obs}$) | Calculated Intensity ($I_{clac}$) |
|---|---|---|---|---|---|---|---|---|---|
| MnFe$_2$O$_4$ | 1 | 1 | 1 | 1 | 8 | 4.861413 | 0.102705 | 8.1 | 7.3 |
| | 2 | 2 | 2 | 0 | 12 | 2.976995 | 0.138635 | 22.6 | 21.7 |
| | 3 | 3 | 1 | 1 | 24 | 2.53879 | 0.148486 | 38.8 | 36.5 |
| | 4 | 2 | 2 | 2 | 8 | 2.430707 | 0.150685 | 6.6 | 7 |
| | 5 | 4 | 0 | 0 | 6 | 2.105053 | 0.156895 | 19 | 13.5 |
| | 6 | 3 | 3 | 3 | 8 | 1.620471 | 0.156264 | 2.5 | 2.3 |
| | 7 | 4 | 4 | 0 | 12 | 1.488498 | 0.148942 | 35.8 | 32.2 |
| Cu$_{0.25}$Mn$_{0.75}$Fe$_2$O$_4$ | 1 | 1 | 1 | 1 | 8 | 4.84621 | 0.114235 | 9.9 | 7.9 |
| | 2 | 2 | 2 | 0 | 12 | 2.967685 | 0.145975 | 26.9 | 26.7 |
| | 3 | 3 | 1 | 1 | 24 | 2.530851 | 0.154694 | 90.5 | 83.9 |
| | 4 | 2 | 2 | 2 | 8 | 2.423105 | 0.156778 | 8.3 | 6.1 |
| | 5 | 4 | 0 | 0 | 6 | 2.098471 | 0.162237 | 21.3 | 16.7 |
| | 6 | 3 | 3 | 3 | 8 | 1.615403 | 0.16059 | 2.8 | 2.4 |
| | 7 | 4 | 4 | 0 | 12 | 1.483843 | 0.153124 | 36.7 | 33.4 |

The peak heights of the XRD patterns were measured using the *WinPlotr* software and then the ratios (observed) were calculated using equation 4 for the selected planes, which are presented below in the table 3. The calculated values are from the refined model.

**Table 3:** The calculated cationic distributions of the samples and the ratios of intensities of (220), (440), (400) peaks.

| Sample | A site | B site | $I_{220}/I_{440}$ | | $I_{220}/I_{400}$ | |
|---|---|---|---|---|---|---|
| | | | Obs | Calc | Obs | Calc |
| MFO | Mn$_{0.179}$Fe$_{0.817}$ | Mn$_{0.825}$Fe$_{1.178}$ | 0.6312 | 0.6739 | 1.8 | 1.607 |
| CMFO | Mn$_{0.245}$Fe$_{0.76225}$Cu$_{0.005}$ | Mn$_{0.505}$Fe$_{1.2341}$Cu$_{0.245}$ | 0.7329 | 0.7994 | 1.46 | 1.59 |

The $I_{220}/I_{440}$, $I_{220}/I_{400}$ ratios indicate the accuracy of the theorized cation distribution among the A and B sites.

## 3.3 Calculation of Lattice Parameter using N-R function

The angular positions of the peaks can be used to determine the cell parameter by Nelson-Riley (N-R) method[9]. The cell parameter (a) is determined from every set of corresponding $d_{hkl}$ spacing and h,k,l values for each peak, then plotted against the N-R function as defined below,

$$F_{NR} = \left(\frac{\cos^2\theta}{\sin\theta}\right) + \left(\frac{\cos^2\theta}{\theta}\right) \quad (5)$$

The y-intercept of the linear regression of the cell parameter vs. $F_{NR}$ plot gives the exact value of the cell parameter. Figure 3 shows the determination of the cell parameter via the aforementioned method and the values of a determined via this method are listed in table 4.

**Fig. 3:** Lattice parameter vs. Nelson-Riley function of (i) $MnFe_2O_4$ and (ii) $Cu_{0.25}Mn_{0.75}Fe_2O_4$

## 3.4 Calculation of cell parameter and other structural parameters

The cell parameter can also be calculated theoretically from the cation-anion distances using the formula below, which can be calculated from the cationic compositions measured experimentally by any other means. If $d_{A-O}$ and $d_{B-O}$ are the average cation-anion bond lengths of A and B sites respectively, the relation between lattice parameter, *a* and the bond lengths can be expressed through the following formula,

$$a_{calc} = \frac{8}{9}(\sqrt{3}d_{A-O} + 3d_{B-O}) \quad (6)$$

There is more than one type of atoms occupying the same position. So weighted mean values of the crystal/ionic radii calculated by R. D. Shanon [10] were used to calculate effective values of $d_{A-O}$ and $d_{B-O}$. The composition calculated by Rietveld refinement method was used to determine the ionic distances.

**Table 4:** The cell parameters determined by N-R method, theoretically calculated form ionic distances compared with the cell parameter, oxygen position parameter values calculated with the Rietveld method.

| Sample | $a_{NR}$ (Å) | Ionic distances | | $a_{calc}$ (Å) | $a_{rietveld}$ (Å) | $r_A$ (Å) | $r_B$ (Å) |
|---|---|---|---|---|---|---|---|
| | | $d_{AO}$ (Å) | $d_{BO}$ (Å) | | | | |
| MnFe$_2$O$_4$ | 8.38047 | 2.03726 | 1.96188 | 8.344086958 | 8.377974 | 0.77726 | 0.69282 |
| Mn$_{0.75}$Cu$_{0.25}$Fe$_2$O$_4$ | 8.40921 | 2.05056 | 1.96207 | 8.36307995 | 8.402696 | 0.790555 | 0.6922665 |

The effective radii of the tetrahedral and octahedral sites were again calculated using the following formulae,

$$r_A = C(Mn_A^{+2}) \cdot r(Mn_A^{+2}) + C(Cu_A^{+2}) \cdot r(Cu_A^{+2}) + C(Fe_A^{+3}) \cdot r(Fe_A^{+3}) \quad (7)$$

$$r_B = \frac{1}{2}[C(Mn_B^{+2}) \cdot r(Mn_B^{+2}) + C(Cu_B^{+2}) \cdot r(Cu_B^{+2}) + C(Fe_B^{+3}) \cdot r(Fe_B^{+3})] \quad (8)$$

The crystal radii of $Mn^{2+}$, $Cu^{2+}$, $Fe^{3+}$ cations at A site have coordination number 4 compared to these ions at site B having coordination number 6.

The position of the $O^{2-}$ anion in the face-centered cubic (fcc) lattice is referred to as a quantity called Oxygen position parameter or anion parameter, u. For ideal spinel structures, $u_{ideal}$ is 0.375 [11,12]. To accommodate substituted cations, the oxygen ions reposition themselves, which gives rise to change in the oxygen position parameter. The relation between effective radii $r_A$, $r_B$ and $r_{O^{-2}}$ are as follows, which can be further exploited to calculate the value of $u$.

$$r_{A(Tet)} = \left(u - \frac{1}{4}\right) a\sqrt{3} - r_{O^{-2}} \quad (9)$$

$$r_{B(Oct)} = \left(\frac{5}{8} - u\right) a - r_{O^{-2}} \quad (10)$$

Evidently, the value of $u$ is larger than that of ideal spinel structure, because of the replacement of $Fe^{3+}$ with larger $Mn^{2+}$ ions and larger $Cu^{+2}$ ions.

The values of the tetrahedral and octahedral bond lengths ($d_{AL}$ and $d_{BL}$), tetrahedral edge ($d_{AE}$) and shared and unshared bond lengths ($d_{BE}$ and $d_{UBE}$) can be calculated using the following formulae,

$$d_{AL} = a\sqrt{3}\left(u - \frac{1}{4}\right) \quad (11)$$

$$d_{BL} = a\left[(3u^2) - \left(\frac{11}{4}\right)u + \frac{43}{64}\right]^{1/2} \quad (12)$$

$$d_{AE} = a\sqrt{2}\left(2u - \frac{1}{2}\right) \quad (13)$$

$$d_{BE} = a\sqrt{2}\,(1 - 2u) \quad (14)$$

$$d_{BEU} = a\left[4u^2 - 3u + \frac{11}{16}\right]^{\frac{1}{2}} \quad (15)$$

The hopping lengths, $L_A$ and $L_B$ between the magnetic ions located at A site and B site respectively can be calculated from the equations below,

$$L_A = a\frac{\sqrt{3}}{4} \quad (16)$$

$$L_B = a\frac{\sqrt{2}}{4} \quad (17)$$

The bond lengths, angles and hopping lengths are listed in table 5.

Reduction of $\theta_1$, $\theta_2$ and $\theta_5$ are indicative of strengthening A-A, A-B interaction, while on the other hand increase in $\theta_3$, $\theta_4$ indicates strengthening of B-B interaction. These effects of super-exchange interactions between A-sites and B-sites ultimately decrease magnetization. In this case, it can be seen as the $Cu^{2+}$ was incorporated $\theta_1$, $\theta_2$ and $\theta_5$ increase and the other two angles got reduced. The changes in the lengths are presented in table 5.

**Table 5:** The calculated values of the ionic distances and the bond angles along with the ideal values of the bond angles [13].

| x | $d_{AL}$ (Å) | $d_{BL}$ (Å) | $d_{AE}$ (Å) | $d_{BE}$ (Å) | $d_{BUE}$ (Å) | $\theta_1$ (°) | $\theta_2$ (°) | $\theta_3$ (°) | $\theta_4$ (°) | $\theta_5$ (°) |
|---|---|---|---|---|---|---|---|---|---|---|
| Ideal | - | - | - | - | - | 125.150 | 154.560 | 90.000 | 125.033 | 79.633 |
| 0.0 | 1.9223 | 2.0338 | 3.1391 | 2.7850 | 2.9647 | 122.8732 | 142.9196 | 93.5867 | 126.0800 | 73.2277 |
| 0.25 | 1.8636 | 2.0858 | 3.0432 | 2.9182 | 2.9810 | 124.4189 | 150.0871 | 91.2207 | 125.5486 | 77.4939 |

The cation-oxygen distances p,q,r,s, cation–cation distances b, c, d, e, f and the interaction angles were calculated using the relations listed in table 6.

**Table 6:** The formula for the cation-anion distances, cation- cation distances and the bond angles [14].

| Cation-anion distances $M$-$O$ (Å) | Cation- cation distances $M$-$M$ (Å) | Bond Angles $\theta$ (°) |
|---|---|---|
| $p = a\left(\dfrac{5}{8} - u\right)$ | $b = \dfrac{\sqrt{2}}{4}a$ | $\theta_1 = \left(\dfrac{p^2 + q^2 - c^2}{2pq}\right)$ |
| $q = a\sqrt{3}\left(u - \dfrac{1}{4}\right)$ | $c = \dfrac{\sqrt{11}}{8}a$ | $\theta_2 = \left(\dfrac{p^2 + r^2 - e^2}{2pr}\right)$ |
| $r = a\sqrt{11}\left(u - \dfrac{1}{4}\right)$ | $d = \dfrac{\sqrt{3}}{4}a$ | $\theta_3 = \left(\dfrac{2p^2 - b^2}{2p^2}\right)$ |
| $s = a\sqrt{3}\left(\dfrac{u}{3} + \dfrac{1}{8}\right)$ | $e = \dfrac{3\sqrt{3}}{8}a$ | $\theta_4 = \left(\dfrac{p^2 + s^2 - f^2}{2ps}\right)$ |
|  | $f = \dfrac{\sqrt{6}}{4}a$ | $\theta_5 = \left(\dfrac{r^2 + q^2 - d^2}{2rq}\right)$ |

**Table 7:** Various possible distances between the ions in spinel structure.

| x | p(Å) | q(Å) | r(Å) | s(Å) | b(Å) | c(Å) | d(Å) | e(Å) | f(Å) |
|---|---|---|---|---|---|---|---|---|---|
| $MnFe_2O_4$ | 2.0319 | 1.9223 | 3.6809 | 3.6639 | 2.96206 | 3.47332 | 3.62776 | 5.44165 | 5.13044 |
| $Mn_{0.75}Cu_{0.25}Fe_2O_4$ | 2.0496 | 1.9077 | 3.6529 | 3.6679 | 2.97080 | 3.48357 | 3.63847 | 5.45771 | 5.14557 |

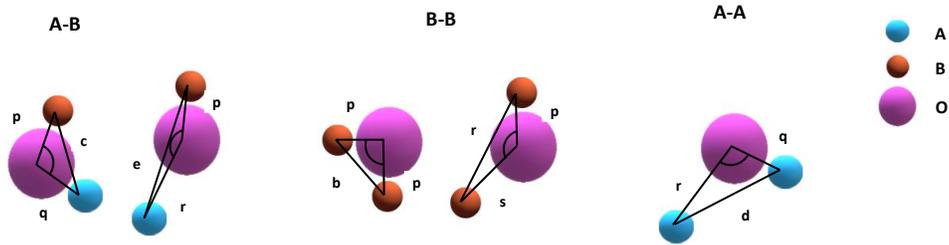

**Fig. 4:** The bond lengths and the interaction angles of an ideal spinel structure. A and B are cations ($Cu^{2+}$, $Mn^{2+}$ or $Fe^{2+/3+}$) represented by blue and brown balls respectively and the oxygen anion ($O^-$) is represented by O.

### *3.5 Neutron Diffraction (ND)analysis, structure refinement and magnetic properties*

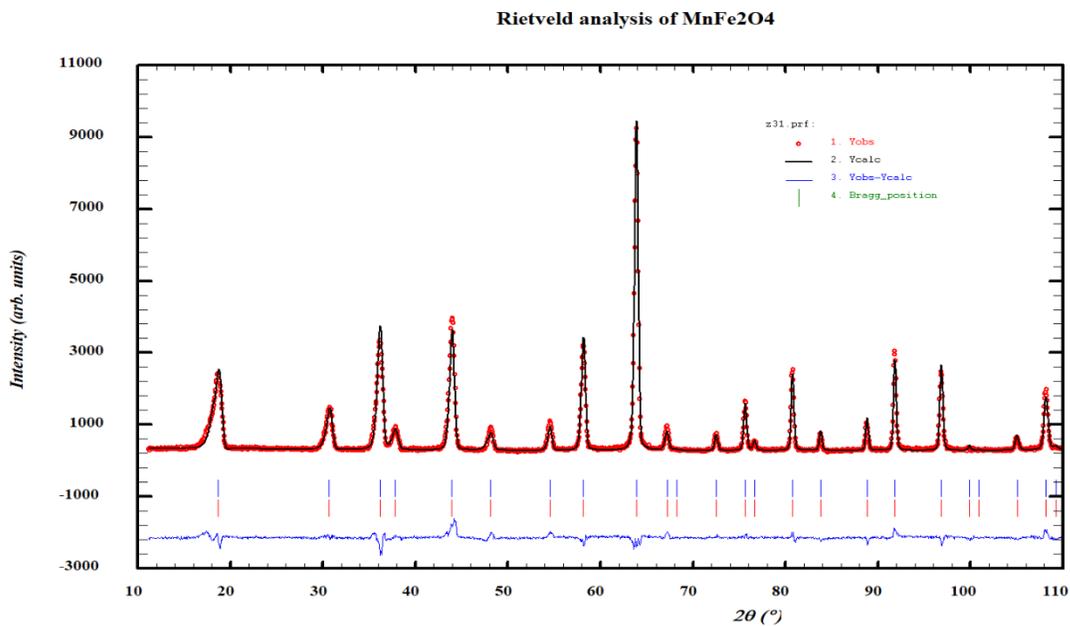

(i)

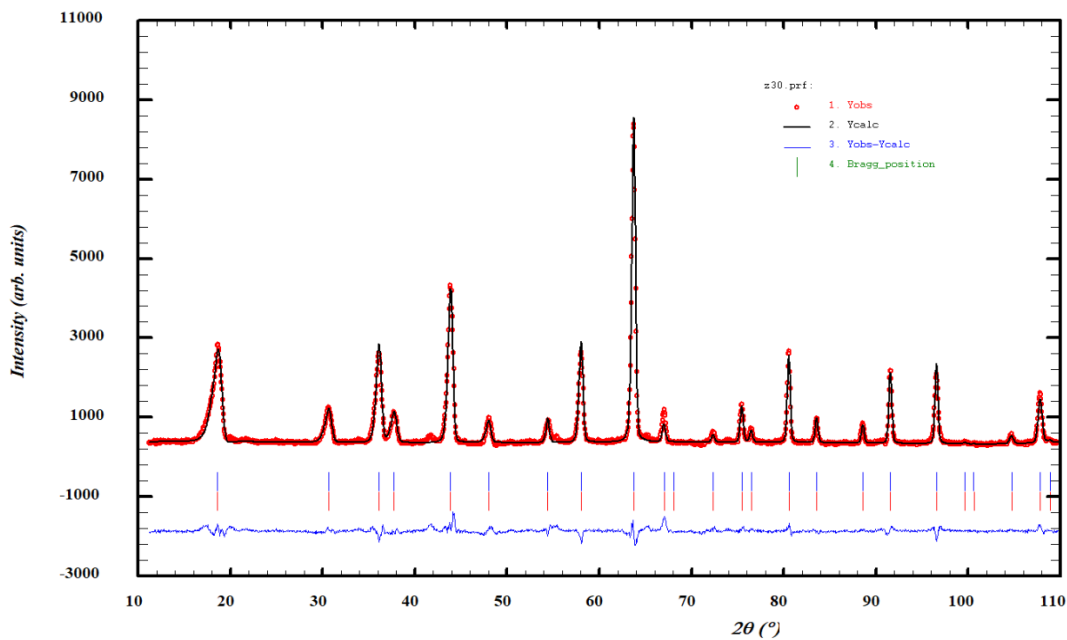

(ii)

**Fig. 5:** The fitted neutron diffraction patterns of (i) Manganese Ferrite ($MnFe_2O_4$) and (ii) Cu doped Manganese Ferrite ($Mn_{0.75}Cu_{0.25}Fe_2O_4$). The red dots represent the observed data and the solid black line represents the calculated pattern by *FullProf* Rietveld refinement software. The first row of blue markers indicates the structural peaks and the red markers in the row below represents the peaks from the magnetic structure.

The magnetic structure of a material may conform with or differ from the molecular structure of the parent material. In this case, from refinement, it was seen that the magnetic peaks (indicated by the red ticks in the fig. 5) coincide with the structural peaks (indicated with the blue ticks). So the structure of both the materials is commensurate and the magnetic propagation vector, *K* is (0,0,0) [15]. The magnetic moments of site A (1/8, 1/8, 1/8) and site B (1/2, 1/2, 1/2) calculated from FullProf code are presented in table 8. The negative sign of the moment of site A indicated that it is antiparallel to the orientation of the B site moment. As all the cations here had a non-zero magnetic moment (Cu, Mn and Fe), Neel's co-linear model was considered appropriate to explain the magnetism of the materials. The results are compared with the magnetic measurements with the VSM[16–18].

**Table 8:** The magnetic moments of A and B sites and net magnetic fields of manganese ferrite and copper doped manganese ferrite samples, calculated from neutron diffraction experiments.

| Cation distribution | Moment of site A, $m_A$ ($\mu_B$) | Moment of site B, $m_B$ ($\mu_B$) | Net magnetic moment, $m_{net} = \lvert m_A - m_B \rvert$ ($\mu_B$) |
| --- | --- | --- | --- |
| $(Mn_{0.179}Fe_{0.817})_A[Mn_{0.825}Fe_{1.178}]_BO_4$ | -5.8305 | 8.6839 | 2.8534 |
| $(Mn_{0.245}Fe_{0.76225}Cu_{0.005})_A[Mn_{0.505}Fe_{1.2341}Cu_{0.245}]_BO_4$ | -5.9242 | 7.60073 | 1.6765 |

### 3.6 The refinement parameters

The measurement of the accuracy of a Rietveld based fitting is represented by the 'Goodness of fit' parameters. The $R_p$, $R_w$, $R_e$, $R_f$, $R_{Bragg}$ and $\chi^2$ are profile factor, weighted profile factor, expected weighted profile factor, crystallographic $R_f$ factor, Bragg's R factor and the reduced Chi-square value respectively [19,20]. The refinement parameters of XRD and neutron diffraction pattern of the samples are listed in table 9 along with the formulae used by the refinement software. The X-ray data has single-phase hence the overall $\chi^2$ values are considerably below 2, which is indicative of good agreement between the model and the XRD pattern. Unlike XRD, the neutron diffraction data has not only structural phase but also the magnetic phase; hence $\chi^2$ value is slightly over 3, in which the structural parameters investigated accords with the X-ray diffraction studies.

**Table 9:** The goodness of fit parameters of the XRD and Neutron diffraction pattern refinements of the samples.

| R-factors | | $MnFe_2O_4$ | | $Mn_{0.75}Cu_{0.25}Fe_2O_4$ | |
|---|---|---|---|---|---|
| | | XRD | Neutron Diffraction | XRD | Neutron Diffraction |
| $R_p$ | $\left[\dfrac{\sum_i \lvert Y_i^{obs} - Y_i^{calc} \rvert}{\sum_i Y_i^{obs}}\right]$ | 59.5 | 13.8 | 54.3 | 15.9 |
| $R_w$ | $\left[\dfrac{\sum_i w_i \lvert Y_i^{obs} - Y_i^{calc} \rvert^2}{\sum_i w_i Y_i^{obs^2}}\right]^{\frac{1}{2}}$ | 45.9 | 13.4 | 48.4 | 15.6 |
| $R_e$ | $\left[\dfrac{N - P + C}{\sum_i w_i Y_i^{obs^2}}\right]^{\frac{1}{2}}$ | 43.3 | 7.4 | 44.6 | 7.89 |
| $R_f$ | $\dfrac{\sum_{hkl} \lvert \lvert F_{hkl}^{in} \rvert^2 - \lvert F_{hkl}^{out} \rvert^2 \rvert}{\sum_{hkl} \lvert F_{hkl}^{in} \rvert^2}$ | 13.7 | 3.72 | 11.6 | 3.34 |
| $R_{Bragg}$ | $\left[\dfrac{\sum_k \lvert I_k^{obs} - I_k^{calc} \rvert}{\sum_k I_k^{obs}}\right]$ | 15.3 | 5.79 | 12.6 | 5.08 |
| $\chi^2$ | $\left(\dfrac{R_{wP}}{R_E}\right)^2$ | 1.122 | 3.28 | 1.18 | 3.893 |

### 3.7 Magnetic measurements using VSM

The magnetic properties of spinel ferrites largely depend on the cation distribution (the distributions of the substitutes in A or B position), which can be varied by synthesis method, annealing temperature, reactants etc. The magnetic hysteresis loops (M-H) are given in figure 6.

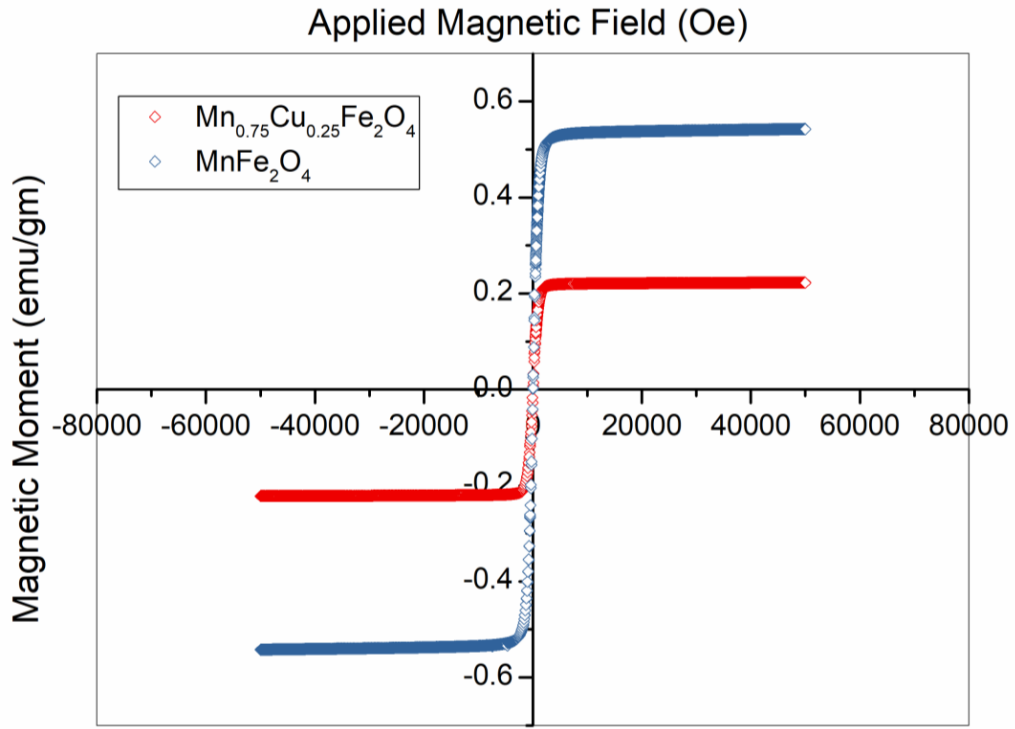

**Fig. 6:** Magnetic hysteresis, *M(H)* curves of (a) Manganese Ferrite (MnFe$_2$O$_4$) and (b) Cu doped Manganese Ferrite (Mn$_{0.75}$Cu$_{0.25}$Fe$_2$O$_4$) represented with blue circles and red squares respectively at T$_a$=300K.

The magnetic moment per unit formula can be calculated using the following formula,

$$\eta = \frac{M_w M_s}{5585} \qquad (18)$$

Here, $M_w$ is the molecular weight of the sample and $M_s$ is the saturation magnetization.

**Table 10:** The Magnetic properties of the samples.

| Sample | Saturation Magnetization, M$_s$ (*emu/gm*) | Remnant Magnetization, M$_r$ (*emu/gm*) | M$_r$/M$_s$ | Coercive field, H$_c$ (*Oersted*) | Magnetic Moment per unit formula, $\eta$ |
|---|---|---|---|---|---|
| Mn$_{0.75}$Cu$_{0.25}$Fe$_2$O$_4$ | 54.24861 | 3.873 | 0.07139 | 31.825 | 2.261032 |
| MnFe$_2$O$_4$ | 31.77113 | 2.574 | 0.08101 | 17.7528 | 1.311956 |

The values of the magnetic properties like Saturation magnetization (M$_s$), Remnant magnetization (M$_r$), their ratio, Coercive field (H$_c$), Magnetic moment per unit formula ($\eta$) etc. were in accordance with the values investigated by L. A. Kafshgari et. al. and M. Khaleghi et. al.

[21,22], the difference in the values can be explained from the different methods of synthesis and annealing temperature. In the afore mentioned works, the samples were synthesized via sol-gel, co-precipitation or hydrothermal method, though solid-state method is exclusively for bulk synthesis and widely used for its feasibility and quality of yielded sample.

### 3.8 Scanning Electron Microscopy (SEM) studies

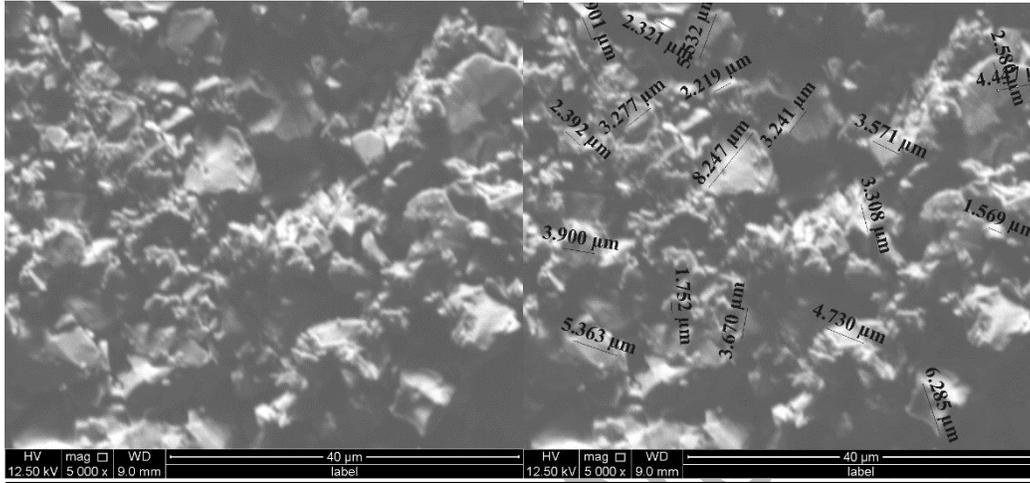

(i)        SEM image of $MnFe_2O_4$

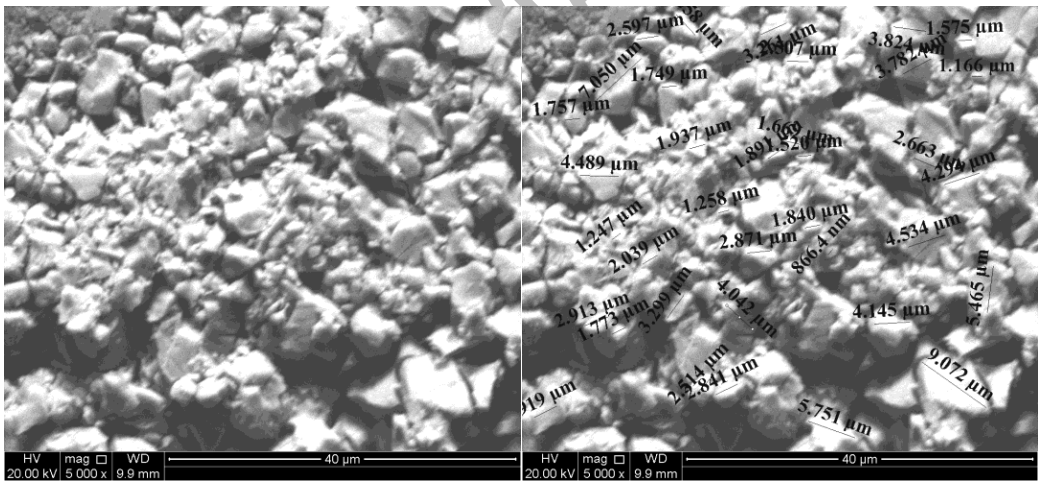

(ii)        SEM image of $Mn_{0.75}Cu_{0.25}Fe_2O_4$

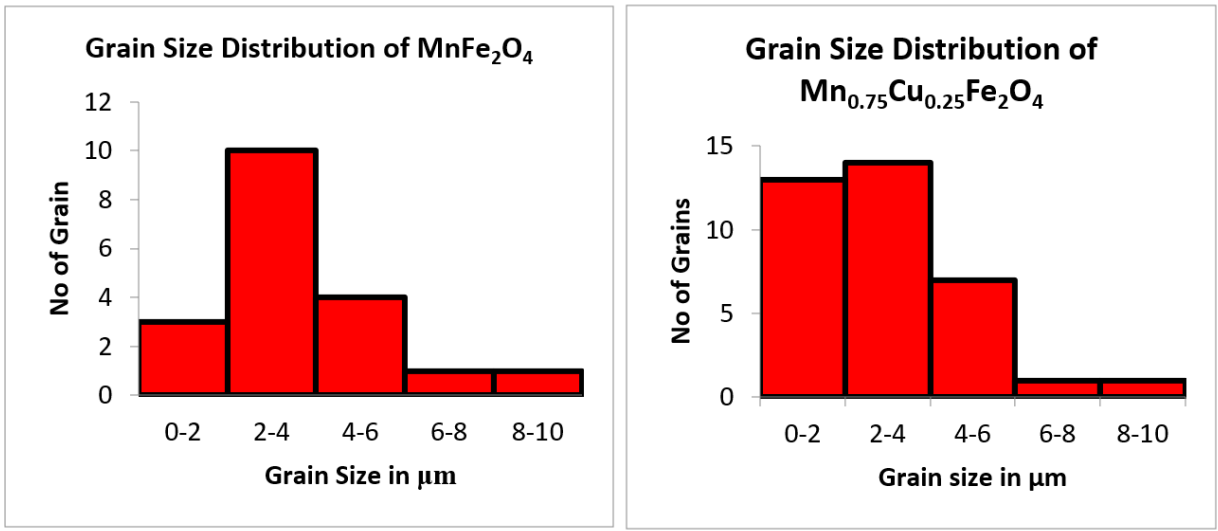

(iii-a)  (iii-b)

**Fig.7:** The scanning electron microscope images of the samples. (i) SEM image of $MnFe_2O_4$, (ii) SEM image of $Mn_{0.75}Cu_{0.25}Fe_2O_4$, (iii) size distributions of the samples.

The SEM images revealed the morphology of the soft manganese ferrite samples. It showed that the sample surface had large grain size with broken edges. It also revealed nonporous and rough surface, which remained the same for both doped and non-doped material. But the morphology of the materials changed with the increment of Cu concentration[23]. The major change could be observed regarding grain size. It could be seen from figure (iii) that the doped sample had more grains in the range of 0-2 micrometer and 2-4 micrometer, while the undoped material had lower number of grains in 0-2 micrometer region but higher number of grains in 2-4 micrometer region.

## 4. Conclusion

In this study, $MnFe_2O_4$ and Cu-doped $MnFe_2O_4$ were synthesized in the conventional solid-state sintering method. The XRD analysis confirms the single-phase formation of the samples, and the surface morphology study showed proper densification of the particles. Furthermore, various parameters were studied and compared by the Rietveld refinement of the XRD pattern and neutron diffraction analysis to observe the effect of doping in the samples. The observed net magnetic moment declined in both as $Mn^{2+}$ ion has larger magnetic moment compared to $Cu^{2+}$. Though, saturation magnetization ($M_s$) and remnant magnetization ($M_r$) both were reduced, their ratio ($M_s/M_r$) increased. The SEM micrographs indicates the abundance of particles having size 0-2 μm and 4-6 μm in Cu doped $MnFe_2O_4$ compared to $MnFe_2O_4$.

## Acknowledgements

The research group is thankful to Centre for Research Reactor (CRR) of Bangladesh Atomic Energy Commission (BAEC) for the operation of research reactor during the experiments, Dr. Sheikh Manjura Hoque, Head & CSO of Materials Science Division, AECD, BAEC for SEM and VSM facilities.